\documentstyle[11pt,aaspp4]{article}

\slugcomment{To appear in ApJ.}

\lefthead{Rots et al.}
\righthead{RXTE Pulsar Absolute Timing Results}

\begin{document}

\title{RXTE Absolute Timing Results for the Pulsars B1821$-$24 and B1509$-$58}

\author{A. H. Rots\altaffilmark{1},\altaffilmark{2}, K. Jahoda, and D. J. Macomb\altaffilmark{1}}
\affil{Laboratory for High Energy Astrophysics, Code 660, NASA, Goddard Space
Flight Center, Greenbelt, MD 20771}
\authoremail{arots@head-cfa.harvard.edu}

\author{N. Kawai}
\affil{Institute of Physical and Chemical Research, Wako (RIKEN),
Saitama 351-01, Japan}
\author{Y. Saito}
\affil{Institute of Space and Astronautical Science,
Sagamihara, Kanagawa 229, Japan}
\author{V. M. Kaspi\altaffilmark{3}}
\affil{Department of Physics and Center for Space Research,
Massachusetts Institute of Technology, Cambridge, MA 02139}
\author{A. G. Lyne} 
\affil{Department of Physics, University of Manchester, Jodrell Bank,
Macclesfield, SK11 9DL, UK}
\author{R. N. Manchester}
\affil{Australia Telescope National Facility, CSIRO, P.O.\ Box 76,
Epping, NSW 2121, Australia}
\author{D. C. Backer and A. L. Somer}
\affil{Astronomy Department and Radio Astronomy Laboratory, University
of California, Berkeley, CA 94720}
\author{D. Marsden and R. E. Rothschild}
\affil{Center for Astrophysics and Space Science, University of California
at San Diego, San Diego, CA 92093}
\altaffiltext{1}{Universities Space Research Association}
\altaffiltext{2}{Present address: Harvard-Smithsonian Center for
Astrophysics, MS81, 60 Garden Street, Cambridge, MA 02138}
\altaffiltext{3}{Hubble Fellow}

\begin{abstract}
Observations with the Rossi X-ray Timing Explorer and the Jodrell
Bank, Parkes, and Green Bank telescopes have enabled us to determine
the time delay between radio and X-ray pulses in the two isolated
pulsars B1821$-$24 and B1509$-$58.  For the former we find that the
narrow X-ray and radio pulse components are close to being coincident
in time, with the radio peak leading by 0.02 period ($60 \pm20 \mu$s),
while the wide X-ray pulse component lags the last of the two wider
radio components by about 0.08 period.  For the latter pulsar we find,
using the standard value for the dispersion measure, that the X-ray
pulse lags the radio by about 0.27 period, with no evidence for any
energy-dependence in the range 2-100~keV.  However, uncertainties in
the history of the dispersion measure for this pulsar make a
comparison with previous results difficult.  It is clear that there
are no perceptable variations in either the lag or the dispersion
measure at time scales of a year or less.
\end{abstract}

\section{Introduction}

In the past, various attempts at absolute timing of pulsar signals
have been made, trying to establish the phase lag between radio and
X-ray pulses.  Examples include the papers on PSR~B1509$-$58 by
\cite{kaw91}, and \cite{ulm93}.  From these we know that there is a
phase lag and that it is, roughly speaking, between 0.25 and 0.35.
First, \cite{buc78} and, more recently, \cite{kan94} have measured the
phase lag between gamma-ray and radio emission for the Vela pulsar.
\cite{mas94} reported for the Crab pulsar that the radio pulse appears
to lag the gamma-ray pulse by about 0.5 ms, based on Figaro II
observations.  This was not confirmed by the CGRO observations of
\cite{nol93} and \cite{ulm94} (respectively, EGRET and OSSE data),
though the former did not comment on the issue.  The latter find the
phase lag to be less than 30~$\mu$s, but with an uncertainty of
300~$\mu$s.  Hence, we feel that it is fair to say
that in the case of the Crab pulsar, for most past high energy
astrophysics space missions, the lag may be considered zero for all
practical purposes because of uncertainties in absolute time keeping.
If there is a lag, it has to be significantly less than 1~ms.

This situation is different for the Rossi X-ray Timing Explorer (RXTE
or XTE) when using one or both of its main instruments, the
Proportional Counter Array (PCA) and the High-Energy X-ray Timing
Experiment (HEXTE).  The precision and accuracy of the RXTE absolute
timing will, in principle, allow measuring phase lags as short as
10~$\mu$s.  This means that for all known X-ray pulsars the X-ray time
keeping will no longer be the dominant source of uncertainty.
However, in most, if not all, cases we will not be able to achieve an
accuracy of 10~$\mu$s since we will be limited by the accuracy of the
radio observations, the differing shapes of the light curves (as well
as variations therein depending on waveband and time), and counting
statistics.

This type of measurement provides important information for a better
understanding of the emission processes involved.  In isolated neutron
stars, powered by spin-down rather than accretion, the pulsed emission
is likely to originate from either a polar cap (\cite{dau82}) or
synchrotron processes in the outer magnetospheric gaps (\cite{che86}).
For more quantitative treatment of the latter type of models see,
e.g., \cite{smi86,rom95}.  Accurate absolute timing data will increase
our knowledge of the precise location where the emission originates
and of the geometry of the magnetic field lines.

This paper presents absolute timing results for two stable radio
pulsars, PSR~B1821$-$24 and PSR~B1509$-$58, using RXTE and the Jodrell
Bank, Parkes, and Green Bank radio telescopes.  For the former pulsar
we will also make a comparison with ASCA observations.  Together with
similar results for the Crab pulsar (PSR~B0531$+$21) which are
described in a separate paper (\cite{rot97}), the experiment
demonstrates the capabilities of the RXTE in this area.  In addition
to the scientific results, it is our intention to provide with this
paper a reference for RXTE absolute timing issues.

\section{RXTE Timing Accuracy}

The RXTE Mission Operations Center (MOC) performs about ten
calibration observations of the spacecraft clock per day.  The
calibration, using the USCCS (\cite{GSFC91}) technique, relies on a
round trip signal, tagged by the spacecraft clock's time stamp, to
determine the spacecraft clock offset.  The method claims to provide
absolute time with an accuracy of 5~$\mu$s.  Parabolic fits to
datasets extending over periods of about a week show deviations of no
more than 1~$\mu$s which is consistent with quoted uncertainties for
the various steps in the procedure.  However, the true uncertainty is
dominated by the clock at the ground station at White Sands which is
only ``required to agree with UTC at the Naval Observatory to within
5000~ns (5~$\mu$s)'', though it ``is typically kept within [\ldots]
2~$\mu$s of UTC'' (\cite{GSFC91}).  Consequently, we shall assume that
RXTE absolute time is correct to within 5~$\mu$s.  In addition,
however, most of the data used in the course of this investigation
suffer from a slight additional degradation, leading us to adopt a
value of 8~$\mu$s for the uncertainty in absolute time for
observations made before April 29, 1997 (MJD 50567).  This degradation
was caused by a known error in the clock calibration ground software
which was corrected on that date.

We have validated these calibrations by astronomical observations of
burst source 1744$-$28 (comparing CGRO-BATSE and RXTE-PCA data) and
the pulsars PSR~B1509$-$58 (period 150~ms) and PSR~B0531$+$21 (Crab,
period 33~ms) to an accuracy of 1~ms.  Beyond that level of accuracy,
uncertainties in the knowledge of the properties of the celestial
objects preclude further validation.  However, the validation
justifies, in our minds, accepting the MOC's calibration results.

All XTE data items are tagged with a time stamp taken from the
spacecraft clock, with a maximum resolution of 1~$\mu$s for PCA data,
8~$\mu$s for HEXTE data.  Measurements of timing delays internal in
the spacecraft were made before launch, using a muon source.  These
measurements compared the true time of the events with the time tags
attached by the instrument data systems and hence include detector, as
well as data system, delays.  The timing delay for PCA events was
determined to range from 16~$\mu$s for most events to 20~$\mu$s at low
pulse heights which is close to expected values.  The delay is the sum
of all analog processing and transfer times, and the time tag is
applied by the Experiment Data System (EDS).  The HEXTE time stamp, on
the other hand, is applied by the HEXTE Data System at the moment the
lower level discriminator is exceeded; the resulting HEXTE instrumental
delay is less than 1~$\mu$s.

The time, as recorded in RXTE FITS files is TT (Terrestrial Time),
with an accuracy of better than 100~$\mu$s.  Additional corrections,
based on the MOC's calibrations can reduce the error to 8~$\mu$s.  The
issue of time systems is dealt with in more detail in the RXTE-GOF WWW
pages\footnote{http://rxte.gsfc.nasa.gov} (see: ``Time
Tutorial''\footnote{http://heasarc.gsfc.nasa.gov/docs/xte/abc/time\_tutorial.html}
and ``Absolute Time
Calibration''\footnote{http://heasarc.gsfc.nasa.gov/docs/xte/abc/time.html}),
and in full detail by \cite{sei92}.

All other sources of timing error that affect barycenter corrections,
are small compared to the clock uncertainties quoted above.  The
$3 \sigma$ errors for the solar system ephemeris and the RXTE orbit
ephemeris are 0.1~$\mu$s and 0.25~$\mu$s, respectively.  In summary,
RXTE time stamps, with and without barycenter correction, can be
corrected to within 5~$\mu$s (8~$\mu$s before MJD 50567) of absolute
time.  We adopted corrections for the instrumental delays of 16~$\mu$s
and 1~$\mu$s for PCA and HEXTE, respectively.

In this paper we present absolute timing results for two stable radio
pulsars, PSR~B1821$-$24 and PSR~B1509$-$58.  We do not include our
extremely high signal-to-noise observations of the Crab pulsar here;
interpretation of those data requires a careful treatment of the
intrinsic timing noise in that source, and will be reported separately
(\cite{rot97}).

\section{Analysis of RXTE Observations \label{sec3}}

We report here on observations made with the two main instruments on
RXTE.  The Proportional Counter Array (PCA) consists of five
Xenon-filled detectors that cover the energy range 2-50 keV with a
combined nominal collecting area of 7000~cm$^2$.  The High-Energy
X-ray Timing Experiment (HEXTE) consists of two clusters of four
NaI/CsI ``phoswich'' scintillation detectors which are usually rocking
to provide a background estimate through a beam switching technique;
the combined nominal collecting area is 800~cm$^2$ per cluster.  Since
we are interested in pulsed signals, we switched off the rocking.

Public RXTE observations of PSR~B1821$-$24 were made on September~16,
1996 (MJD~50342) in the context of an RXTE-ASCA clock
cross-calibration project.  The total RXTE exposure time was 6559~s,
divided over two orbits.  The PCA data configuration used was {\em
GoodXenon} which records all good events detected in the Xenon chamber
with full timing accuracy of 1~$\mu$s.  To improve the signal-to-noise
ratio, we only used the photons detected in the top Xenon layer.  The
pulsed signal is too weak to be detected by HEXTE in such a short
exposure.  In this paper we will also present the ASCA observations
that were made contemporaneously, as part of the cross-calibration
program.  These observations are not the same as presented by
\cite{sai97}, but are described in more detail by Saito et al.\ (1997b).

For PSR~B1509$-$58, we present proprietary observations made with the
PCA in the same {\em GoodXenon} data configuration, and HEXTE
observations using an event data configuration with 8~$\mu$s time
resolution.  These observations cover the period January through
October 1996.

The observations were analyzed using the program faseBin, developed by
one of us (AHR) and publicly distributed as part of the Ftools
analysis package by the GSFC HEASARC.  The program selects only good
events, calculates the absolute pulse phase, based upon the radio
pulsar timing ephemeris, and bins the photon events in a
two-dimensional histogram of energy channel {\em vs} pulse phase.

The time from the FITS files is corrected to 8~$\mu$s accuracy and
transformed from TT to TDB (Barycentric Dynamical Time), using the
RXTE orbit ephemeris and the JPL DE-200 solar system ephemeris (see
\cite{sta82,sta90}).  Both of these ephemerides are accurate to better
than 1~$\mu$s.  The uncertainty introduced by the TDB$-$TT term is
approximately 2~$\mu$s.  In summary, we feel confident that our time
stamps are correct to 10~$\mu$s.  The same code and ephemeris are used
to transform the timing ephemeris to TDB.

The two-dimensional histograms are then collapsed over one or more
energy ranges to create light curves as a function of absolute phase,
or over one or more phase ranges to allow phase-resolved
spectroscopy.

All X-ray pulse phases in this paper refer to the peak of the pulse.
The timing ephemerides used are given in Table~\ref{tbl1}.

\placetable{tbl1}

In the near future, we intend to switch the faseBin program from the
DE-200 to the DE-405 ephemeris.  At this place, for future reference,
we shall briefly deal with the issues involved in that change-over.
Aside from higher accuracy, there are two fundamental changes and two
possible sources of inconsistencies.  The two fundamental changes are:
\begin{enumerate}
\item Although both ephemerides are based on the epoch J2000.0, DE-200
uses the older FK-5 reference system while DE-405 is tied to the
International Celestial Reference System (ICRS), as adopted by the IAU
at its 1997 General Assembly in Kyoto, Japan.
\item Due to more accurate masses of the solar system bodies, the
position of the barycenter has shifted.  The difference can amount to
several milliseconds.
\end{enumerate}
In tying X-ray observations to radio timing ephemerides, the two
potential inconsistencies, as it turns out, do in many cases not cause
a problem:
\begin{enumerate}
\item The radio timing ephemerides, till the present, have been
derived using DE-200.  Fortunately, this cancels out since the time of
arrival of the pulse, or zero-phase, (column 5 in Table~\ref{tbl1}) is
given in (i.e., converted back to) geocentric time, but only to the
level of tens of micro-seconds.  Hence, it cannot be ignored for
milli-second pulsars.  In the future, radio ephemerides will be based
on DE-405, and an code will be provided indicating which planetary
ephemeris has been used.
\item The spacecraft orbit ephemeris may be (and usually is) provided
in J2000.0 geocentric coordinates referenced to FK-5.  The error
incurred by adding such vectors to J2000.0 ICRS vectors is
proportional to to the length of the geocentric FK-5 vector.  Since
the misalignment between the two reference systems is about
20~milli-arcseconds, the maximum error arising from this misalignment
is approximately 2~ns times the distance of the spacecraft from the
geocenter, expressed in earth radii.  Hence, the error is definitely
negligible, not only for spacecraft in low earth orbit like RXTE, but
also for more elliptical orbits like AXAF.
\end{enumerate}.

\section{PSR~B1821$-$24}

\subsection{Comparison with Jodrell Bank and ASCA Observations}

\begin{figure}
\epsscale{0.5}
\plotone{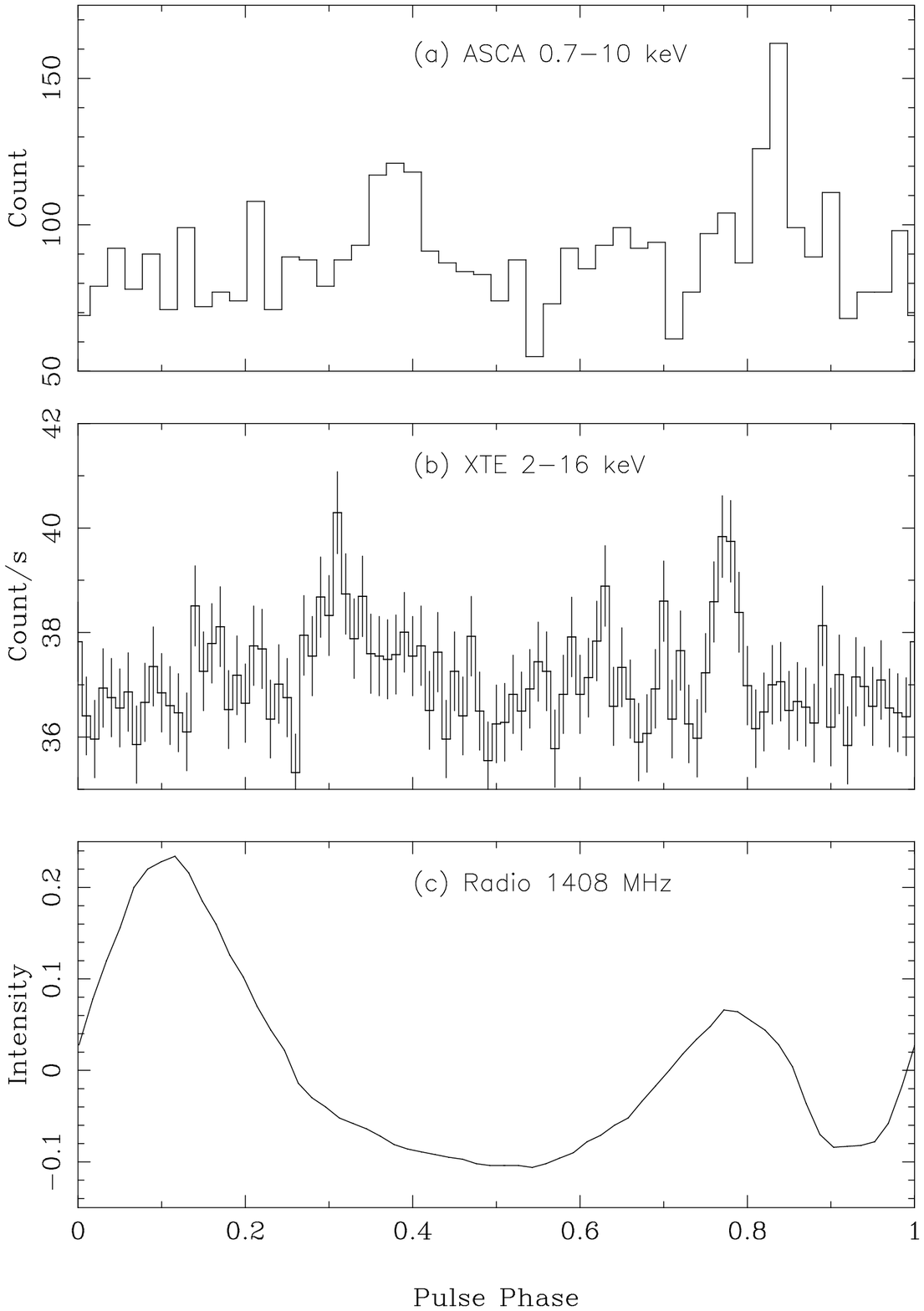}
\caption{PSR~B1821$-$24.  Light curves based on (a)~ASCA GIS
(0.7-10~keV), (b)~RXTE PCA (2-16~keV), (c)~Jodrell Bank radio
(1408~MHz) observations, using timing ephemeris B1821$-$24J.  The ASCA
light curve is labeled in total counts per bin (at 48 bins per
period), the RXTE light curve in counts per second per bin (at 100
bins per period) with $1 \sigma$ errors indicated.  The instrumental
broadening in the radio profile amounts to about 0.1 in pulse
phase. Note that the zero point of the phase of this ephemeris used in
this figure is different from the one used in Fig.~\ref{fig1a}.\label{fig1}}
\end{figure}

The timing ephemeris was derived from observations made with the
Jodrell Bank Mark~IA radio telescope, at an observing frequency of
1408~MHz, assuming a dispersion measure of 119.86~pc~cm$^{-3}$.  The
pulse profile, showing two pulse components is included in
Fig.~\ref{fig1}.  In order to be consistent with the nomenclature in
\cite{bac97}, we shall refer to the pulse component at phase 0.1 as
number 2, the one at phase 0.8 as number 1.  The time resolution of
the observations was about 300~$\mu$s, or one-tenth of a period.  The
error in the determination of the dispersion measure leads to an
uncertainty of about 150~$\mu$s, or about 0.05 in phase, at infinite
frequency.  The timing ephemeris is labeled in Table~\ref{tbl1} as
B1821$-$24J.  Note that the zero point of the phase of this ephemeris
is different from that of the Green Bank ephemeris.

The RXTE signal-to-noise ratio is such that we only present one light
curve: PCA 2-16~keV (channel~5-49).  There are two clear peaks, a
narrow one at phase~0.8 and an asymmetric one at phase~0.3, consistent
with the separation between the peaks seen in the ASCA result reported
by \cite{sai97} but slightly offset in phase.  The narrow pulse
component has a width less then 0.04 in phase, corresponding to
100~$\mu$s.  Analysis of the two orbits separately shows that these
two peaks are clearly present in both sets while all other features do
not consistently appear in both; the latter must therefore be either
spurious or time-variable and will not be considered here.

We present the RXTE light curve in Fig.~\ref{fig1}, together with the
ASCA GIS and radio light curves.  The background level (internal and
cosmic background, and unpulsed source) is included.  The total pulsed
flux is about 1\% of the combined unpulsed flux and backgrounds.
These data prove that it is possible to detect with RXTE a {\em
narrow} pulsed signal with a flux of only 0.5\% of the unpulsed flux
in a single orbit observation.

When the original ASCA observations were reported on by \cite{sai97},
it was not at that time possible to compare the ASCA and radio phases
in an absolute sense since the uncertainty in the ASCA clock was
approximately 1~ms.  This problem has been solved for the ASCA
observations obtained in the context of the RXTE-ASCA clock
cross-calibration project that are presented in this paper.  The
uncertainty in the ASCA clock has been reduced to 200~$\mu$s (see:
\cite{sai97b}\footnote{Also:
http://heasarc.gsfc.nasa.gov/docs/asca/newsletters/gis\_time\_assign5.html})
and all phases in Fig.~\ref{fig1} are calculated as absolute phases,
applying all known clock corrections.  The ASCA light curve exhibits
the same features as the RXTE one, albeit that they are shifted by
approximately 0.05 period in phase ($\approx 150~\mu$s).  Errors of
this magnitude have been observed in ASCA's clock and attributed to
drift due to temperature variations.  Hence, we feel justified in
assuming that RXTE's absolute timing is the more correct one.

The phase information is summarized in Table~\ref{tbl2}.  Note that
this table uses the radio pulse component numbering from \cite{bac97}.


\placefigure{fig1}
\placetable{tbl2}

\subsection{Comparison with Green Bank Observations}

Recently, \cite{bac97} published a new radio pulse profile of
PSR~B1821$-$24 based on observations made with the NRAO\footnote{The
National Radio Astronomy Observatory is a facility of the National
Science Foundation operated under cooperative agreement by Associated
Universities, Inc.}  42~m telescope at Green Bank, WV, at observing
frequencies of 800 and 1395~MHz.  Owing to the use of the Spectral
Processor backend, the temporal resolution of the pulse profile is
much higher that that of the Jodrell Bank one, revealing the existence
of a third radio pulse component.  \cite{bac97} compared the result
with ASCA observations, but no absolute phase comparison could be
made.  We have analyzed the RXTE observation using the timing
ephemeris that was derived from the Green Bank observations (see
Table~\ref{tbl1}, B1821$-$24G).  The dispersion measure was, of
course, determined for each observing session separately, based on the
dual frequency data.

\begin{figure}
\epsscale{0.5}
\plotfiddle{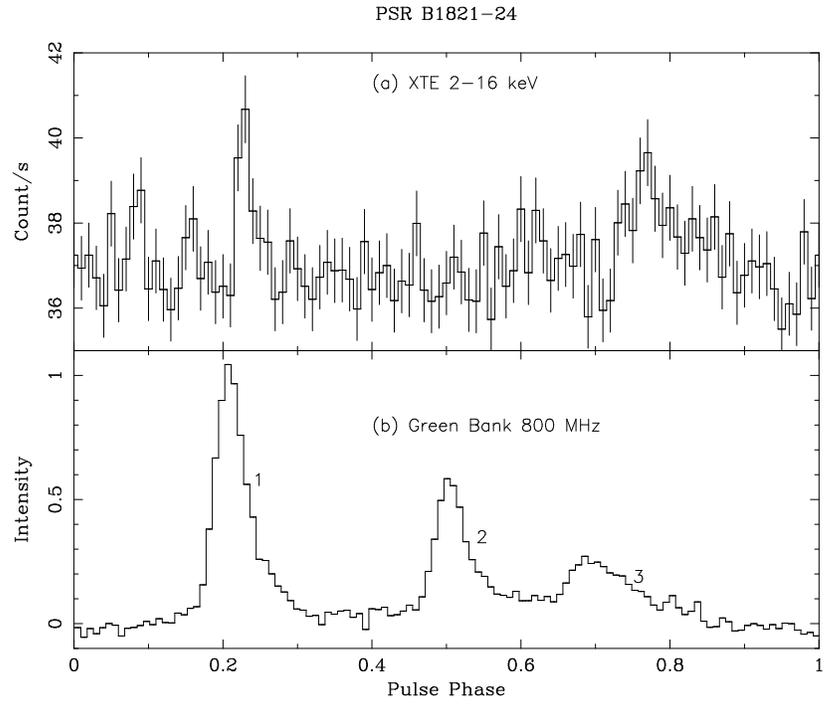}{3.7in}{-90.0}{50}{50}{-150}{350}
\caption{PSR~B1821$-$24.  Light curves based on (a)~RXTE PCA
(2-16~keV), and (b)~Green Bank radio (800~MHz) observations, using
timing ephemeris B1821$-$24G.  The RXTE light curve is labeled in
counts per second per bin (at 100 bins per period) with $1 \sigma$
errors indicated. The radio pulse components are identified using the
same numbering as Backer and Sallmen (1997). \label{fig1a}}
\end{figure}

\placefigure{fig1a}

The resulting light curves, presented in Fig.~\ref{fig1a}, are quite
convincing.  There is good agreement between the narrow X-ray pulse
component and radio pulse component~1, as correctly anticipated by
\cite{bac97}, but the broad X-ray pulse component lags radio
component~3 by about 0.08~period.  The X-ray version of pulse
component~1 seems to lag the radio component by 60~$\mu$s which is
not consistent with their being temporally coincident.  The
accuracy of the X-ray phase, 10~$\mu$s (see Section \ref{sec3}), is
dominated by the uncertainty in the RXTE clock; the accuracy of the
radio phase is approximately 10~$\mu$s ($1 \sigma$), mainly due to
uncertainty in the dispersion measure.  We have summarized the phases
of the pulse component peaks in Table~\ref{tbl2}.  The quoted errors
are approximate $1 \sigma$ values.  One should note that the
difference between the pulse profiles derived from the Jodrell Bank
and Green Bank observations, as shown in Figs.~\ref{fig1} and
\ref{fig1a} is due in part to the difference in temporal resolution,
but also in part to the difference in observing frequency.  The
spectral characteristics of the two components are very different at
radio frequencies: at higher frequencies pulse component~2 is
generally stronger than pulse component~1.  The high resolution pulse
profile at 1395~MHz presented by \cite{bac97} provides an easier
comparison.

It is interesting to note that, after the Crab pulsar, we have here,
once again a pulsar where the main radio and X-ray pulse component
peaks are coincident to better than 20~milli-periods.

This public RXTE observation only yielded 6559~s of exposure.  Two of us
(YS and NK) have obtained a proprietary observation of much longer
duration.  It seems prudent, therefore, to defer discussion of other
characteristics, such as spectral properties, to the publication of
those observations.

\section{PSR~B1509$-$58}

The timing data for PSR~B1509$-$58 were obtained using the Parkes
telescope\footnote{The Parkes telescope is part of the Australia
Telescope which is funded by the Commonwealth of Australia for
operation as a National Facility by CSIRO.} in a continuation of the
observational program described by \cite{kas94}; the observing
frequency was 1400~MHz, the assumed dispersion measure
253.2~pc~cm$^{-3}$.  The radio pulse profile displays a single pulse
component with a full width at half power of about 0.1~period; the
phase is referenced to the peak of that pulse with an accuracy of
about 1~ms.  The timing ephemeris is provided in Table~\ref{tbl1}.

\subsection{Timing Analysis}

In X-rays, this pulsar has been monitored during the entire RXTE
mission to date.  The monitoring observations each lasted at least
2000~s and were done approximately once a month.  Fig.~\ref{fig2}
shows the light curves in the 2-16~keV PCA band for ten of these
observations.  The vertical line is drawn at phase 0.27.  There are
clearly variations in these light curves, but they are consistent with
a constant (X-ray~$-$~radio) peak phase lag of $0.27 \pm0.01$~period,
as indicated by the distribution, in phase, of the peak bins of the
ten light curves.  We have also analyzed the cross-correlation
functions between the ten light curves.  The centroids are all
distributed within 0.008 of the average.  From this we conclude that
the variation in phase is no more than 0.005~period, comparable to the
uncertainty in the radio timing ephemeris.  As a matter of fact, there
appear to be systematic variations at this level that are correlated
with the different timing ephemeris entries.  Our phase lag result is
consistent with the phase lag derived by \cite{kaw91}, but is
different from that determined by \cite{ulm93}.  We will return to
this issue in the discussion section, below.  The shapes of the light
curves agree with those obtained by both previous investigations.

\placefigure{fig2}

Fig.~\ref{fig3} shows the accumulated light curves for the bands 2-4,
4-8, 8-16, 16-32, 32-64, 64-128~keV (the first four PCA, the last
three HEXTE).  These data are consistent with a phase lag of 0.27,
independent of energy.  Note that the changes in pulse profile with
energy are probably not significant, as indicated by the spectral
analysis, below.

\placefigure{fig3}

\subsection{Spectral Analysis}

We performed phase-resolved spectral analysis, using the
program XSPEC.  For this purpose we accumulated PCA and HEXTE spectra
in eight slices of 0.05~period, from phase 0.20 through phase 0.55.
The average count rate spectrum in the phase range from 0.7 through
1.1 was used as the value for the unpulsed components of internal and
cosmic background, as well as the source itself, and subtracted from
spectra of the pulsed radiation.  It transpired that the spectra could
all be fit well with a simple power-law model with interstellar
absorption (\cite{mor83}).  On the basis of fits to spectra near the
peak of the pulse, we adopted a value for $N_H$ of $6 \times 10^{22}$.
Hence, only photon spectral index and total flux were allowed to vary.
The somewhat surprising result is that all spectra are consistent with
a single value for the photon index: $1.345 \pm0.010$; i.e., the
photon flux density (photons cm$^{-2}$ s$^{-1}$ keV$^{-1}$) is
proportional to $E^{-1.345\pm0.01}$.  The derived photon index values,
with formal errors and reduced $\chi^2$, are listed in
Table~\ref{tbl3}.  Fig.~\ref{fig4} presents the spectral fit for
phase~0.25.  It is illustrative of the fits at other pulse phases.

\placetable{tbl3}

The value for the photon index of 1.345 is consistent with the one
obtained by \cite{kaw93}.  They found $1.30 \pm0.05$.  \cite{mat94}
determined the index to be $1.68 \pm0.09$ for the energy range 50
keV~-~5 MeV, based on OSSE observations and speculated that there has
to be a break in the spectrum between 20 and 80~keV.  A broken
power-law fit to our spectra does not improve the fit.  Hence, we
conclude that the break is most likely to occur above 50~keV.  A more
comprehensive spectral analysis has been presented by (\cite{mar97}).

The surprise lies in the fact that there is no significant change of
photon index with phase.  \cite{kaw91} have speculated that there may
be two components making up PSR~B1509$-$58's X-ray pulse shape.  The
light curves seem to support this notion.  But if such is the case,
then the two components are, at least to RXTE, spectrally
indistinguishable.

\placefigure{fig4}

\subsection{Discussion}

The significance of our results for the difference in phase lag
between, on the one hand, this paper and \cite{kaw91}, and, on the
other, \cite{ulm93} is considerable.  \cite{kaw91} find a value of
$0.25 \pm0.02$ for the 2-11~keV band, which is consistent with the
value we find of $0.27 \pm0.01$ for the 2-16~keV band.  This phase lag
refers explicitly to the peak.  \cite{ulm93} find (fitting to the
shape of the Kawai et al.\ 1991 light curve) a phase lag of $0.32 \pm
0.02$ for BATSE data covering the band 20-400~keV (the OSSE
observation is less relevant in this context).  We should also
mention, at this point, the results from the balloon experiment
``Welcome'', reported by \cite{gun94}.  Covering the range 94-240~keV,
these observations are consistent, both spectrally and temporily, with
the Ginga data; the uncertainties preclude a stronger statement.  The
inconsistency with the \cite{ulm93} result (8~ms) is especially troubling
because the CGRO absolute time information is qualitatively much
better than that of the other missions.
It cannot be summarily dismissed as a CGRO clock or software
error since such would have had very noticeable effects on, for
instance, the Crab pulsar light curves from CGRO instruments.

We attempted to resolve the issue by reprocessing five years of BATSE
observations of PSR~B1509$-$58 from the public CGRO archive, using all
data at energies higher than 32~keV.  The result is shown in
Fig.~\ref{fig5}.  Cross-correlation analysis of the pulse profiles
reveals that the BATSE light curve is shifted by 0.03 period, or 5~ms.
Although this difference is smaller than for \cite{ulm93}'s result, it
still exceeds the acceptable bounds, it is still too large to be
unnoticed in the Crab data, and it raises the question why this BATSE
time lag is different from what was found previously.

\placefigure{fig5}

There are only two
explanations possible that can reconcile the results of the three
space-based investigations (Kawai et al.\ 1991, Ulmer et al.\ 1993,
and the present paper): the phase lag varies with energy; or the phase
lag varies in time.

Energy dependence seemed unlikely, judging from the full energy range
of \cite{ulm93}'s BATSE and OSSE observations, but remained possible.
Our data show convincingly that there is no variation of phase lag
over the range 2-100~keV.  Even if the reader may not be persuaded by
Fig.~\ref{fig3} that this statement applies above 50~keV, the
hypothesis requires a considerable hardening of the spectrum at phases
above 0.30; this is clearly ruled out by our spectral analysis.  This
is corroborated by the combination of Ginga and Welcome observations
by \cite{kaw91} and \cite{gun94} which make it seem very unlikely that
a break in the spectrum would occur below 100~keV.

\begin{figure}
\epsscale{0.5}
\plotone{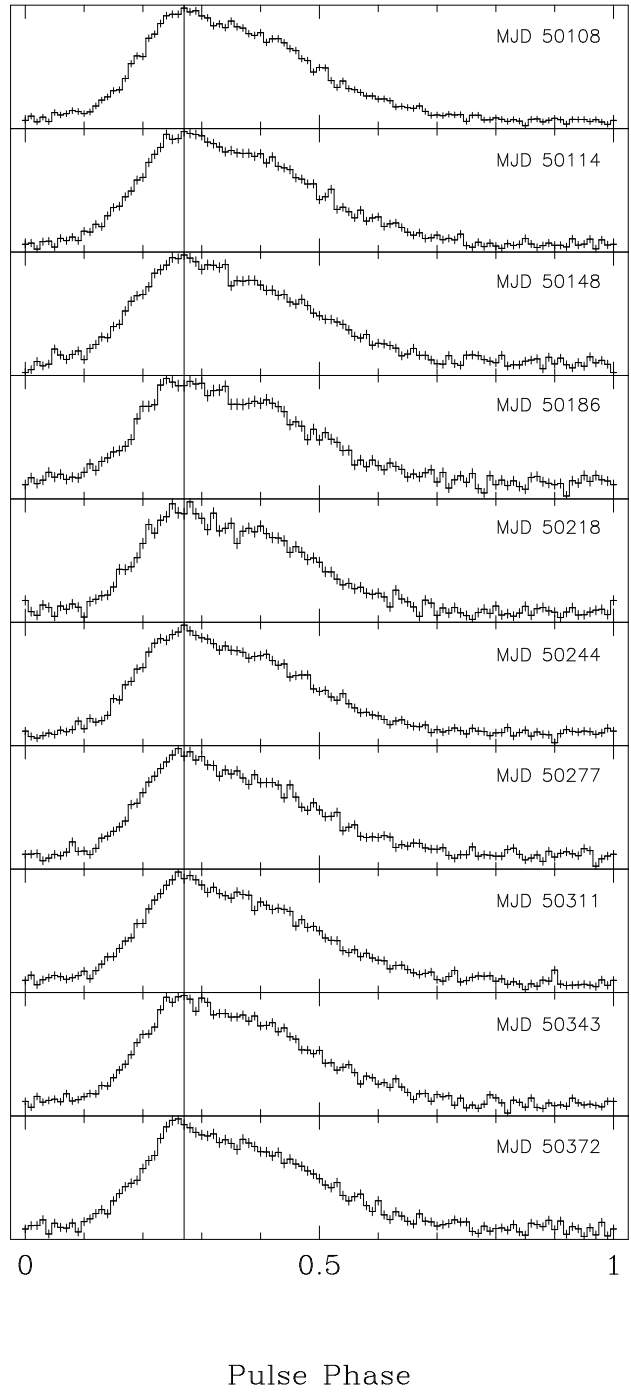}
\caption{PSR~B1509$-$58.  Light curves for ten epochs (PCA 2-16~keV).
The error bars represent $1 \sigma$ errors.  The vertical line
indicates phase 0.27.  \label{fig2}}
\end{figure}

\begin{figure}
\epsscale{0.5}
\plotone{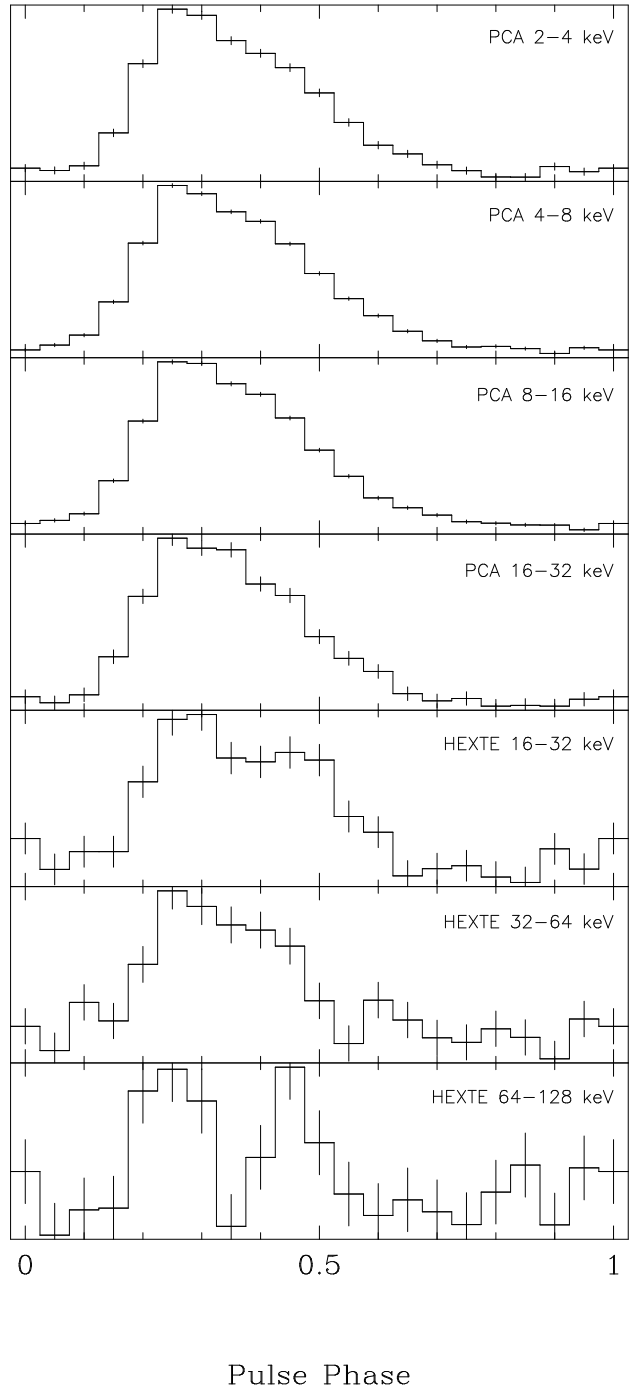}
\caption{PSR~B1509$-$58.  Accumulated light curves for seven energy
bands.  The vertical scale is arbitrary; the error bars represent
$1 \sigma$ errors.  \label{fig3}}
\end{figure}

\begin{figure}
\epsscale{0.5}
\plotfiddle{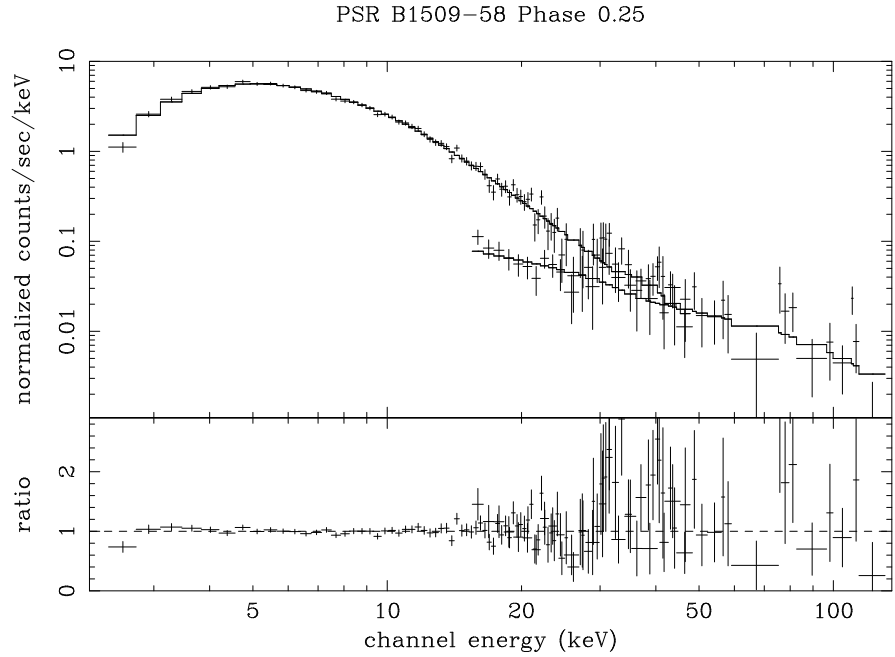}{3.7in}{-90.0}{50}{50}{-200}{350}
\caption{PSR B1509$-$58.  Spectral fit for pulse phase~0.25.  The
crosses ($1 \sigma$ error bars) in the upper panel represent the
observations, the solid lines the model convolved with the response
matrices.  The left hand segment pertains to PCA data, the right hand
segment to HEXTE data.  The lower panel presents the ratio $data /
model$.  \label{fig4}}
\end{figure}

\begin{figure}
\epsscale{0.5}
\plotone{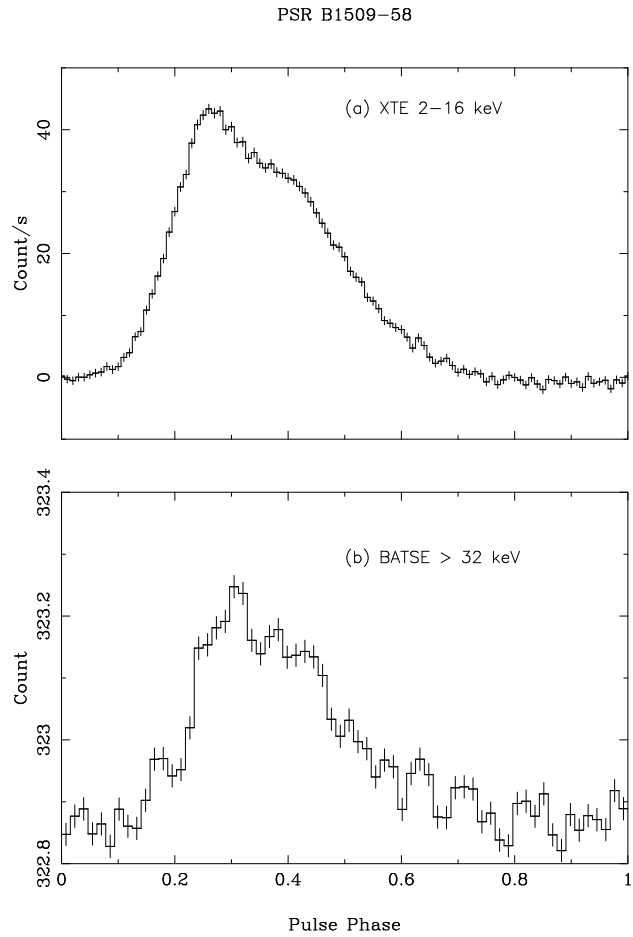}
\caption{PSR~B1509$-$58.  Accumulated light curves based on (a)~RXTE
PCA (2-16~keV) and (b)~BATSE ($>$~32~keV) observations.  The error bars
represent $1 \sigma$ errors.  \label{fig5}}
\end{figure}

Time evolution of the phase lag was all but ruled out by \cite{ulm93},
on the basis of their BATSE and OSSE observations.  This is
corroborated by our data, at least on time scales of one year or less
(Fig.~\ref{fig2}).  It is also worth mentioning that the radio
ephemerides were created over short durations so that phase drift due
to low level timing noise observed in the pulsar cannot be the cause
of the discrepancy.

A change in the pulsar dispersion measure would result in an apparent
change in phase offset between radio and X-ray energies.  However, the
dispersion measure change required to explain the discrepancy is over
an order of magnitude larger than is expected based on similar results
for most other pulsars (\cite{bac93}, and references therein); it
requires an increase of more than 0.5~pc~cm$^{-3}$ per year.  In our
analysis we have used the value for the dispersion measure of
$253.2\pm1.9$~pc~cm$^{-3}$ given by \cite{kas94}.  This same value was
used by all investigators quoted in the present paper.  Recently
(MJD~50780-50784), one of us (RNM) has made a new measurement of the
dispersion measure and obtained a value of $255.3\pm0.3$~pc~cm$^{-3}$
($2 \sigma$ error).  A preliminary analysis of nearly contemporaneous
RXTE observations, using this new value, yields a phase offset of
$0.29\pm0.01$.  At first glance, this appears to resolve the
inconsistency, and it very well may, but one has to be extremely
careful: the two determinations of the dispersion measure are still
marginally consistent with each other and there is no irrefutable
evidence that the dispersion measure actually changed.  If we just
take the extremes as indicated by the quoted errors, the change in
dispersion measure over seven years may have been anywhere between 0
and 4~pc~cm$^{-3}$.

In conclusion, we note that a gradual change in dispersion measure
from 253.2~pc~cm$^{-3}$ in 1990 to 255.3~pc~cm$^{-3}$ in 1997 would
explain the changes in phase lag.  But the plausibility of such a
scenario needs to be confirmed by future monitoring of the dispersion
measure.

\acknowledgements

Support for one of us (VMK) was provided by Hubble Fellowship grant
number HF-1061.01-94A from the Space Telescope Science Institute,
which is operated by the Association of Universities for Research in
Astronomy, Inc., under NASA contract NAS5-26555.  We are grateful to
an anonymous referee for helpful comments.

\clearpage

\clearpage

\begin{table*}
\scriptsize
\begin{center}
\caption{Radio Timing Ephemerides \label{tbl1}}
\hspace*{-1cm}
\begin{tabular}{llllllllll}
\multicolumn {1} {c} {Pulsar} & 
\multicolumn {1} {c} {RA} &
\multicolumn {1} {c} {Dec} &
\multicolumn {1} {c} {MJD Range} &
\multicolumn {1} {c} {$t_{0,geo}$} &
\multicolumn {1} {c} {$\nu$} &
\multicolumn {1} {c} {$\dot{\nu}$} &
\multicolumn {1} {c} {$\ddot{\nu}$} &
\multicolumn {1} {c} {Rms\tablenotemark{a}} \\
\multicolumn {1} {c} { } & 
\multicolumn {1} {c} {(J2000.0)} &
\multicolumn {1} {c} {(J2000.0)} &
\multicolumn {1} {c} { } &
\multicolumn {1} {c} {(MJD(UTC))} &
\multicolumn {1} {c} {$({\rm s}^{-1}$)} &
\multicolumn {1} {c} {$(10^{-12}{\rm s}^{-2})$} &
\multicolumn {1} {c} {$(10^{-24} {\rm s}^{-3})$} &
\multicolumn {1} {c} { } \\
\tableline
B1821$-$24J\tablenotemark{b} & $18^{\rm h} 24^{\rm m} 32.008^{\rm s}$ & $-$24\arcdeg 52\arcmin 11.12\arcsec &
50059 - 50372 & 50215.000000023 & 327.4056597973296 &
\phn -0.173520 & \phn \phn \phn 0.0 & 22.2 \nl

B1821$-$24G\tablenotemark{c} & $18^{\rm h} 24^{\rm m} 32.008^{\rm s}$ & $-$24\arcdeg 52\arcmin 10.70\arcsec &
47826 - 50660 & 49243.000000025 & 327.4056743697863 &
\phn -0.173521 & \phn \phn \phn 0.0581 & \phn 6.8 \nl

B1509$-$58 & $15^{\rm h} 13^{\rm m} 55.627^{\rm s}$ & $-$59\arcdeg \phn 8\arcmin \phn 9.54\arcsec &
49730 - 50117 & 49923.000000072 & \phn \phn 6.6284166684779 &
-67.4303 & 1990 & \phn 8.8 \nl

B1509$-$58 & $15^{\rm h} 13^{\rm m} 55.627^{\rm s}$ & $-$59\arcdeg \phn 8\arcmin \phn 9.54\arcsec &
50114 - 50296 & 50205.000000764 & \phn \phn 6.6267743270631 &
-67.3824 & 1950 & \phn 9.1 \nl

B1509$-$58 & $15^{\rm h} 13^{\rm m} 55.627^{\rm s}$ & $-$59\arcdeg \phn 8\arcmin \phn 9.54\arcsec &
50242 - 50462 & 50352.000000582 & \phn \phn 6.6259186740744 &
-67.3579 & 1950 & \phn 8.0 \nl


\tablenotetext{a}{In milli-periods}
\tablenotetext{b}{Based on Jodrell Bank observations}
\tablenotetext{c}{Based on Green Bank observations}

\end{tabular}
\end{center}
\end{table*}

\normalsize

\begin{table*}
\begin{center}
\caption{PSR~B1821$-$24 Pulse Peak Phases \label{tbl2}}
\begin{tabular}{llccc}

Spectral band & Timing ephemeris & Pulse 1 & Pulse 2 & Pulse 3 \\
\tableline
X-ray (2-16 keV)               & B1821$-$24J & 0.775 &       & 0.31 \\
Radio (Jodrell Bank, 1408 MHz) & B1821$-$24J & 0.785 & 0.105 & \\
X-ray (2-16 keV)               & B1821$-$24G & $0.229 \pm 0.003$  &
       & $0.77 \pm 0.01$ \\
Radio (Green Bank, 800 MHz)    & B1821$-$24G & $0.209 \pm 0.003$  &
 $0.50 \pm 0.01$  & $0.69 \pm 0.01$ \\
\end{tabular}
\end{center}
\end{table*}

\begin{table*}
\begin{center}
\caption{PSR~B1509$-$58 Fitted Photon Index as a Function of Phase \label{tbl3}}
\begin{tabular}{ccccc}

Phase &Net Count Rate &Photon Index & Error   & Reduced $\chi^{2}$ \\
\tableline
0.20  &     29.2      &   1.389     & 0.027   &   0.909 \\
0.25  &     45.2      &   1.356     & 0.018   &   1.097 \\
0.30  &     43.4      &   1.341     & 0.018   &   1.050 \\
0.35  &     37.6      &   1.343     & 0.021   &   1.021 \\
0.40  &     34.5      &   1.342     & 0.022   &   1.081 \\
0.45  &     29.0      &   1.341     & 0.026   &   0.861 \\
0.50  &     20.3      &   1.419     & 0.037   &   1.261 \\
0.55  &     12.8      &   1.397     & 0.056   &   0.838 \\
\end{tabular}
\end{center}
\end{table*}

\end{document}